\documentstyle[12pt]{article}
\begin{document}
\begin{titlepage}
\begin{flushright}
IC/2001/62\\
hep-th/0106240
\end{flushright}
\vspace{10 mm}

\begin{center}
{\Large Localization of Bulk Form Fields on Dilatonic Domain Walls}

\vspace{5mm}

\end{center}
\vspace{5 mm}

\begin{center}
{\large Donam Youm\footnote{E-mail: youmd@ictp.trieste.it}}

\vspace{3mm}

ICTP, Strada Costiera 11, 34014 Trieste, Italy

\end{center}

\vspace{1cm}

\begin{center}
{\large Abstract}
\end{center}

\noindent

We study the localization properties of bulk form potentials on dilatonic 
domain walls.  We find that bulk form potentials of any ranks can be 
localized as form potentials of the same ranks or one lower ranks, for 
any values of the dilaton coupling parameter.  For large enough values of 
the dilaton coupling parameter, bulk form potentials of any ranks can 
be localized as form potentials of both the same ranks and one lower ranks.  

\vspace{1cm}
\begin{flushleft}
June, 2001
\end{flushleft}
\end{titlepage}
\newpage

The Randall-Sundrum (RS) scenario \cite{rs1,rs2} provides an alternative 
compactification method, where our four-dimensional spacetime is realized 
as a 3-brane on which the bulk graviton can be localized even with noncompact 
extra spatial dimensions due to warped spacetime.  Localization of various 
bulk fields on the brane has been studied, e.g., Refs. 
\cite{gw,pom,bg,grn,chn,iod,kho,tac}.  In particular, it was shown that, 
whereas the bulk scalar can be localized on the RS domain wall, bulk photon 
and form potentials cannot be localized \cite{pom,kss}.  Later, it was found 
out \cite{dl} that the bulk three-form potential, which is Hodge-dual to 
bulk scalar in five-dimensional bulk spacetime, can also be localized as a 
two-form potential in one lower dimensions with a choice of a modified 
Kaluza-Klein (KK) zero mode ansatz.  (Cf. See also Ref. \cite{sil}.)  It was 
proposed in Ref. \cite{pl}, whose work was extended to the supersymmetric 
case in Ref. \cite{dls}, that a $U(1)$ field on the brane is originated 
rather from two bulk two-form potentials.  Alternative methods for localizing 
bulk $U(1)$ field on the brane through topological Higgs mechanism \cite{oda} 
and by adding a potential of the bulk $U(1)$ field to the brane action 
\cite{gn} were also proposed.  

We showed \cite{youm1,youm2,youm3} that dilatonic domain walls can localize 
bulk gravity, provided that the tension of the wall is positive.  Bulk fields 
with various spins, including bulk $U(1)$ field, were shown \cite{youm2} 
to be localized on such dilatonic domain walls, in the sense that the 
KK zero modes of the bulk fields are normalizable.  It is the purpose of 
this paper to study localization of bulk form potentials of various 
ranks, which were not considered in our previous work.  Unlike the 
case of non-dilatonic RS domain wall, any bulk $p$-form potentials can 
be localized on the dilatonic domain wall both as $p$-form potentials and 
as $(p-1)$-form potentials in one lower dimensions, provided the dilaton 
coupling parameter $a$ is large enough.  Furthermore, for any values of $a$, 
any bulk $p$-form potential can be localized on the dilatonic wall as a 
$p$-form potential or as a $(p-1)$-form potential in one lower dimensions.  

We begin by summarizing dilatonic domain wall solution and localization of 
bulk graviton, studied in Refs. \cite{youm1,youm2,youm3}.  
The total action for the dilatonic domain wall solution is the sum of the 
$D$-dimensional action in the bulk of the domain wall: 
\begin{equation}
S_{\rm bulk}={1\over{2\kappa^2_D}}\int d^Dx\sqrt{-G}\left[{\cal R}-{4\over
{D-2}}\partial_M\phi\partial^M\phi-e^{-2a\phi}\Lambda\right],
\label{bulkact}
\end{equation}
and the $(D-1)$-dimensional action on the domain wall world volume:
\begin{equation}
S_{\rm DW}=-\int d^{D-1}x\sqrt{-\gamma}\sigma_{DW}e^{-a\phi},
\label{dwact}
\end{equation}
where $\gamma$ is the determinant of the induced metric $\gamma_{\mu\nu}
=\partial_{\mu}X^M\partial_{\nu}X^NG_{MN}$ on the domain wall worldvolume, 
$\sigma_{DW}$ is the energy density (or tension) of the domain wall, 
$M,N=0,1,...,D-1$ and $\mu,\nu=0,1,...,D-2$.  In this paper, we consider 
the case in which the gravity can be localized on the domain wall.  
For such case, the domain wall solution is ${\bf Z}_2$-symmetric and 
has naked singularities on both sides of the wall \cite{youm1}:
\begin{eqnarray}
G_{MN}dx^Mdx^N&=&{\cal W}\left[-dt^2+dx^2_1+\dots+ds^2_{D-2}\right]+dy^2,
\cr
\phi&=&{1\over a}\ln(1-K|y|),\ \ \ \ \ \ \ \ \ \ 
{\cal W}=(1-K|y|)^{8\over{(D-2)^2a^2}},
\cr
K&=&{{(D-2)a^2}\over 2}\sqrt{\Lambda\over{2\Delta}},\ \ \ \ \ \ 
\Delta\equiv{{(D-2)a^2}\over 2}-2{{D-1}\over{D-2}},
\label{dildwsol}
\end{eqnarray}
the domain wall tension, which can be fixed by the boundary condition at 
$y=0$, has the following fine-tuned positive value:
\begin{equation}
\sigma_{\rm DW}={1\over\kappa^2_D}{{8K}\over{(D-2)a^2}}=
{4\over\kappa^2_D}\sqrt{\Lambda\over{2\Delta}},
\label{dwtens}
\end{equation}
and the effective $(D-1)$-dimensional gravitational constant has the 
nonzero value $\kappa^2_{D-1}={{\Delta+4}\over 2}\sqrt{\Lambda\over{2\Delta}}
\kappa^2_D$.  Note, from the expression for $K$ in Eq. (\ref{dildwsol}) we 
see that such solution exists only for $\Lambda<0$ [$\Lambda>0$] when $\Delta
<0$ [$\Delta>0$].  We have shown \cite{youm1,youm2,youm3} that the 
normalizable KK zero mode for the bulk graviton exists and therefore the 
gravity can be localized on such dilatonic domain wall, for any values of $a$. 

We consider $p$-form potentials in the bulk of such dilatonic domain wall.  
The action for a massless $p$-form potential $A_{M_1\dots M_p}$ with the 
field strength $F_{M_1\dots M_{p+1}}=(p+1)\partial_{[M_1}A_{M_2\dots 
M_{p+1}]}$ in the $D$-dimensional bulk spacetime is given by
\begin{equation}
S_p=-{1\over{2\cdot(p+1)!}}\int d^Dx\sqrt{-G}G^{M_1N_1}\cdots G^{M_{p+1}
N_{p+1}}F_{M_1\dots M_{p+1}}F_{N_1\dots N_{p+1}}.
\label{pact}
\end{equation}
If we choose the following KK zero mode ansatz for the $p$-form potential
\begin{equation}
A_{\mu_1\dots\mu_p}(x^{\mu},y)=a_{\mu_1\dots\mu_p}(x^{\mu}),
\label{pformpot}
\end{equation}
then the action (\ref{pact}) becomes
\begin{eqnarray}
S_p&=&-{1\over{2\cdot(p+1)!}}\int^{1/K}_{-1/K}dy\left(1-K|y|\right)^{{4
(D-2p-3)}\over{(D-2)^2a^2}}
\cr
& &\times\int d^{D-1}x\sqrt{-g}g^{\mu_1\nu_1}\cdots g^{\mu_{p+1}\nu_{p+1}}
f_{\mu_1\dots\mu_{p+1}}f_{\nu_1\dots\nu_{p+1}},
\label{pactkk}
\end{eqnarray}
where $g_{\mu\nu}(x^{\rho})$ is the KK zero mode for the bulk metric $G_{MN}
(x^{\mu},y)$ and $f_{\mu_1\dots\mu_{p+1}}\equiv(p+1)\partial_{[\mu_1}a_{\mu_2
\dots\mu_{p+1}]}$ is the field strength of the $(D-1)$-dimensional 
$p$-form potential $a_{\mu_1\dots\mu_p}$.  The bulk $p$-form potential 
$A_{M_1\dots M_p}$ can be localized on the dilatonic domain wall, if the 
$y$-integral in Eq. (\ref{pactkk}) is finite, which is the case when
\begin{equation}
p<{{D-3}\over 2}+{{(D-2)^2a^2}\over 8}={{(D-2)(\Delta+4)}\over 4}.
\label{convcond}
\end{equation}
As observed in Ref. \cite{kss}, non-dilatonic domain wall ($a=0$) therefore 
cannot localize $p$-form potentials with $p\geq(D-3)/2$.  In particular, 
the RS domain wall (the $(D,a)=(5,0)$ case) can localize only a 0-form 
field or a scalar field.  However, the dilatonic domain walls ($a\neq 0$) 
can additionally localize the higher rank $p$-form potentials, as long as 
the dilaton coupling parameter $a$ is large enough:  The higher the rank 
$p$ of the form potential, the larger the value of $a$ required for 
localizing the form potential.  All the bulk form potentials (up to rank 
$D-1$) can be localized on the wall, when  
\begin{equation}
a^2\geq {{4(D+3)}\over{(D-2)^2}}
\ \ \ \ \ \ \ \Leftrightarrow\ \ \ \ \ \ \ 
\Delta\geq {8\over{D-2}}.
\label{lcpcnd1}
\end{equation}
So, for example, the five-dimensional dilatonic domain wall in string 
theories obtained by compactifying branes with one type of constituent 
brane on a Ricci flat manifold, for which $\Delta=4$, can localize bulk 
form potentials of all ranks.  

The problem with the KK zero mode ansatz of the form (\ref{pformpot}) is that 
the ansatz of such form for the pair $A_{\mu_1\dots\mu_p}$ and 
$A_{\mu_1\dots\mu_{D-p-2}}$ are not compatible with the following 
Hodge-duality formula in the bulk spacetime:
\begin{equation}
\sqrt{-G}G^{M_1N_1}\cdots G^{M_{D-p-1}N_{D-p-1}}\tilde{F}_{N_1\dots N_{D-p-1}}
={1\over{(p+1)!}}\epsilon^{M_1\dots M_{D-p-1}N_1\dots N_{p+1}}F_{N_1\dots 
N_{p+1}}.
\label{hodgdual}
\end{equation}
Another problem is that a higher rank form potential hodge-dual to a 
lower rank form potential which can be localized on the wall may not be 
localizable, as can be seen from the criterion (\ref{convcond}).  To resolve 
such contradictions, it was proposed in Ref. \cite{dl} to choose the KK zero 
mode ansatz for a $p$-form potential with $p>(D-1)/2$ as
\begin{equation}
A_{\mu_1\dots\mu_{p-1}y}(x^{\mu},y)={\cal W}^{-{{D-2p-1}\over 2}}(y)
a_{\mu_1\dots\mu_{p-1}}(x^{\mu}),
\label{2ndkkans}
\end{equation}
with all the other components vanishing.  Then, ansatz (\ref{pformpot}) 
and (\ref{2ndkkans}) are compatible with with Eq. (\ref{hodgdual}), and 
the bulk Hodge-duality formula (\ref{hodgdual}) reduces to the 
following Hodge-duality formula in $D-1$ dimensions:
\begin{equation}
\sqrt{-g}g^{\mu_1\nu_1}\cdots g^{\mu_{D-p-1}\nu_{D-p-1}}\tilde{f}_{\nu_1
\dots\nu_{D-p-1}}={1\over{p!}}\epsilon^{\mu_1\dots\mu_{D-p-1}\nu_1\dots
\nu_p}f_{\nu_1\dots\nu_p}.
\label{hdgdual2}
\end{equation}
Substituting the ansatz (\ref{2ndkkans}) into the bulk action (\ref{pact}), 
we obtain 
\begin{eqnarray}
S_p&=&-{1\over{2\cdot p!}}\int^{1/K}_{-1/K}dy(1-K|y|)^{-{{4(D-2p-1)}\over
{(D-2)^2a^2}}}
\cr
& &\times\int d^{D-1}x\sqrt{-g}g^{\mu_1\nu_1}\cdots g^{\mu_p\nu_p}
f_{\mu_1\dots\mu_p}f_{\nu_1\dots\nu_p}.
\label{kkact2}
\end{eqnarray}
From this we see that the criterion for a bulk $p$-form potential with 
the KK zero mode of the form (\ref{2ndkkans}) to be localized on the 
wall is
\begin{equation}
p>{{D-1}\over 2}-{{(D-2)^2a^2}\over 8}=-{{(D-2)\Delta}\over 4}.
\label{kkbdcr}
\end{equation}
So, for any values of $a$, all the bulk $p$-form potentials with $p>(D-1)/2$ 
can be localized on the wall as a $(p-1)$-form potential.  In the bulk 
of dilatonic domain wall ($a\neq 0$), bulk form potentials with the lower 
ranks can be additionally localized on the wall in such a manner:  The 
larger the value of $a$, the less stringent the lower bound on $p$ for 
such localization.  All the bulk $p$-form potentials with $p\geq 1$ can be 
localized on the dilatonic wall as $(p-1)$-form potentials, if
\begin{equation}
a^2\geq{{4(D-1)}\over{(D-2)^2}}
\ \ \ \ \ \ \ \ \Leftrightarrow\ \ \ \ \ \ \ \ \ 
\Delta\geq 0.
\label{lccrt2}
\end{equation}
So, for example, all the domain walls obtained from (intersecting) 
branes (with equal charges) in string theories through the 
compactification on Ricci flat manifolds, for which $\Delta=4/N$ with 
$N\in{\bf Z}_+$, can localize any rank bulk $p$-form potentials as 
$(p-1)$-form potentials in $D-1$ dimensions.  

When the KK zero mode is chosen to be of the form (\ref{pformpot}) 
for the lower rank form potentials and of the form (\ref{2ndkkans}) for 
the higher rank form potentials, the only cases in which the non-dilatonic 
domain walls ($a=0$) cannot localize the bulk form potentials are 
$p=(D-3)/2$ and $p=(D-1)/2$, as can be seen from Eqs. 
(\ref{convcond},\ref{kkbdcr}).  For the RS domain wall ($D=5$), these 
correspond to bulk $U(1)$ field and two-form field.  On the other 
hand, dilatonic domain walls can localize bulk $p$-form potentials 
with $p=(D-3)/2$ and $p=(D-1)/2$ when KK zero modes  are chosen 
to be of the forms (\ref{pformpot}) and (\ref{2ndkkans}), 
respectively, for any values of $a$.  So, bulk $p$-form potentials of 
any ranks can be localized on the dilatonic domain wall (with any $a$) 
as $p$-form potentials or $(p-1)$-form potentials in $D-1$ dimensions.  
Furthermore, from Eqs. (\ref{lcpcnd1},\ref{lccrt2}) we see that bulk 
$p$-form potentials of any ranks can be localized on the dilatonic 
domain walls as both $p$-form potentials and $(p-1)$-form potentials 
in $D-1$ dimensions, if the condition (\ref{lcpcnd1}) is satisfied.  

We now explicitly study the KK modes of massless bulk $p$-form potentials 
by considering the following equations of motion obtained from the bulk 
action (\ref{pact}):
\begin{equation}
{1\over\sqrt{-G}}\partial_{M_1}\left[\sqrt{-G}G^{M_1N_1}\cdots G^{M_{p+1}
N_{p+1}}F_{N_1\dots N_{p+1}}\right]=0.
\label{peqn}
\end{equation}
We will find above all that the KK zero modes of a bulk form potentials 
are indeed given by Eqs. (\ref{pformpot},\ref{2ndkkans}).  

First, we consider the dimensional reduction of a bulk $p$-form potential 
to a form potential of the same rank $p$.  By using the gauge degrees of 
freedom, we can take the gauge conditions $A_{M_1\dots M_{p-1}y}=0$.  We 
consider the following KK mode ansatz for the bulk $p$-form potential:
\begin{equation}
A_{\mu_1\dots\mu_p}(x^{\mu},y)=a^{(m)}_{\mu_1\dots\mu_p}(x^{\mu})u_m(y),
\label{pkkans}
\end{equation}
where $a^{(m)}_{\mu_1\dots\mu_p}$ is assumed to satisfy the following field 
equations for a massive $p$-form potential in $(D-1)$-dimensional flat 
spacetime:
\begin{equation}
\partial^{\mu_1}f^{(m)}_{\mu_1\dots\mu_{p+1}}+m^2a^{(m)}_{\mu_2\dots\mu_{p+1}}
=0,
\label{fpeq}
\end{equation}
along with the gauge conditions $\partial^{\mu_1}a^{(m)}_{\mu_1\dots\mu_p}=0$, 
where $f^{(m)}_{\mu_1\dots\mu_{p+1}}\equiv(p+1)\partial_{[\mu_1}a^{(m)}_{\mu_2
\dots\mu_{p+1}]}$ is the field strength of $a^{(m)}_{\mu_1\dots\mu_p}$.  
Then, the equations of motion (\ref{peqn}) for the bulk $p$-form potential 
with the bulk metric (\ref{dildwsol}) substituted reduce to the following 
form of the Sturm-Liouville equation satisfied by $u_m(y)$:
\begin{equation}
\partial_y\left[{\cal W}^{{D-2p-1}\over 2}\partial_yu_m\right]=
m^2{\cal W}^{{D-2p-3}\over 2}u_m.
\label{sleq}
\end{equation}
The operator ${\cal L}=\partial_y\left({\cal W}^{{D-2p-1}\over 2}\partial_y
\right)$ is self-adjoint, provided the boundary condition 
$\left.[({\cal W}^{{D-2p-1}\over 2}u^{\prime}_n)u_m-({\cal W}^{{D-2p-1}\over 2}
u^{\prime}_m)u_n]\right|^{1/K}_{-1/K}=0$ is satisfied.  For such case, the 
eigenvalue $m^2$ is real and the eigenfunctions $u_m$ with different 
eigenvalues are orthogonal to each other w.r.t. the weighting function 
$w(y)={\cal W}^{{D-2p-3}\over 2}$, i.e., $\int^{1/K}_{-1/K}dy\,u_m(y)u_n(y)
w(y)=0$ for $m^2\neq n^2$.  
By using a new $y$-dependent function $\tilde{u}_m={\cal W}^{{D-2p-1}\over 
4}u_m$, we can bring the Sturm-Liouville equation (\ref{sleq}) into the 
following form of the Schr\"odinger equation with zero energy eigenvalue:
\begin{equation}
-{{d^2\tilde{u}_m}\over{dy^2}}+V(y)\tilde{u}_m=0,
\label{scheq}
\end{equation}
where the potential is given by
\begin{equation}
V(y)=\textstyle{{D-2p-1}\over 4}{\cal W}^{-1}{\cal W}^{\prime\prime}+
\textstyle{{(D-2p-1)(D-2p-5)}\over 16}{\cal W}^{-2}({\cal W}^{\prime})^2+
m^2{\cal W}^{-1}.
\label{potential}
\end{equation}
In the case of the KK zero mode ($m=0$), by substituting the expression for 
the warp factor (\ref{dildwsol}) into Eq. (\ref{potential}), we obtain 
the following explicit expression for the potential:
\begin{equation}
V(y)={{2(D-2p-1)K^2}\over{(D-2)^4a^4}}{{2(D-2p-1)-(D-2)^2a^2}\over{(1-K|y|)^2}}
-{{4(D-2p-1)K}\over{(D-2)^2a^2}}\delta(y).
\label{potexp}
\end{equation}
From this we see that the solution to the Schr\"odinger equation (\ref{scheq}) 
satisfies the boundary condition $\tilde{u}^{\prime}_0(0^+)-
\tilde{u}^{\prime}_0(0^-)=-{{4(D-2p-1)K}\over{(D-2)^2a^2}}\tilde{u}_0(0)$.  
The solution to Eq. (\ref{scheq}) satisfying this boundary condition is 
$\tilde{u}_0(y)\sim(1-K|y|)^{{2(D-2p-1)}\over{(D-2)^2a^2}}$.  So, the KK 
zero mode is constant: $u_0(y)={\cal W}^{-{{D-2p-1}\over 4}}\tilde{u}_0(y)=
{\rm constant}$.  This zero mode $u_0(y)$ is normalizable if its norm 
\begin{equation}
\int^{1/K}_{-1/K}dy\,u^2_0(y)w(y)\sim \left.(1-K|y|)^{{4(D-2p-3)+(D-2)^2a^2}
\over{(D-2)^2a^2}}\right|^{1/K}_{-1/K}
\label{norm}
\end{equation}
is finite, which is the case when $4(D-2p-3)+(D-2)^2a^2>0$.  This 
normalization condition coincides with the condition (\ref{convcond}) 
obtained by considering the effective action (\ref{pactkk}).  For such case, 
the normalized zero mode is given by $u_0(y)=\sqrt{{{4(D-2p-3)+(D-2)^2a^2}
\over{2(D-2)^2a^2}}K}$.  We make couple of comments on the KK zero mode.  
First, had we just considered the Sturm-Liouville equation (\ref{sleq}) with 
$m=0$, the most general form of the KK zero mode would have been given by
\begin{equation}
u_0(y)=c_1+c_2{\cal W}^{{{(D-2)^2a^2}\over 8}-{{D-2p-1}\over 2}},
\label{genkkzero}
\end{equation}
where $c_1$ and $c_2$ are integration constants.  This solution expressed in 
terms of $\tilde{u}_0={\cal W}^{{D-2p-1}\over 4}u_0$ is also a general 
solution to the Schr\"odinger equation (\ref{scheq}) in the region $y\neq 0$.
The boundary condition at $y=0$ due to the $\delta$-function term in the 
potential (\ref{potexp}) requires that $c_2=0$.  However, when just the 
Sturm-Liouville equation (\ref{sleq}) is considered, such restriction due to 
the boundary condition at $y=0$ does not exist.  If we use the general zero 
mode (\ref{genkkzero}), then the bulk action (\ref{pact}) takes the form
\begin{eqnarray}
S_p&=&-{1\over{2\cdot p!}}\int^{1/K}_{-1/K}dy\left(c_1\varpi^{{2(D-2p-3)}
\over{(D-2)^2a^2}}+c_2\varpi^{1-{{2(D-2p+1)}\over{(D-2)^2a^2}}}\right)^2
\int d^{D-1}x\sqrt{-g}f^{(0)}_{\mu_1\dots\mu_{p+1}}f^{(0)\,\mu_1\dots\mu_{p+1}}
\cr
& &+\tilde{c}_2\int^{1/K}_{-1/K}dy\int d^{D-1}x\sqrt{-g}a^{(0)}_{\mu_1\dots
\mu_p}a^{(0)\,\mu_1\dots\mu_p},
\label{geneffact}
\end{eqnarray}
where $\varpi\equiv 1-K|y|$ and the constant $\tilde{c}_2$ is proportional 
to $c_2$.  So, with $c_2\neq 0$, we have finite mass term
\footnote{For the non-dilatonic domain wall case, the integration interval 
for the $y$-integration is infinite, so the mass term would diverge.} 
for the $p$-form potential in the effective action, which is contradictory 
to the fact that $u_0$ is the KK zero mode.  Furthermore, the finiteness 
of the kinetic term in Eq. (\ref{geneffact}) requires larger value of $a$ 
than the value satisfying Eq. (\ref{lcpcnd1}).  Second, we have seen 
that the KK zero modes for the bulk $p$-form potentials are independent of 
the extra spatial coordinate $y$.  Although such zero modes are normalizable, 
one might argue that the zero modes are not localized on the wall because 
they are spread evenly in the bulk (rather than localized sharply near the 
wall).  On the other hand, the KK zero mode of the bulk scalar field, which 
is widely regarded as being localized on the wall, is also independent of 
$y$.   Furthermore, the KK zero mode of the bulk graviton is also independent 
of $y$, if the bulk graviton $h_{\mu\nu}$ is defined as ${\cal W}
(\eta_{\mu\nu}+h_{\mu\nu})dx^{\mu}dx^{\nu}+dy^2$.  Perhaps, we might wish to 
choose to consider the product $u^2_0(y)w(y)$ for determining distribution 
of the bulk field across the $y$-direction rather than the zero mode $u_0$ 
itself, since it is this product that appears as the integrand of the $y$ 
integral in the kinetic term of the effective action and may be interpreted 
as being related to the form potential charge density along the 
$y$-direction.  Then, we find that the bulk $p$-form potential is localized 
around $y=0$, spread evenly in the bulk and localized around $|y|=1/K$, 
when $p<(D-3)/2$, $p=(D-3)/2$ and $p>(D-3)/2$, respectively.  So, for $D=5$, 
the distribution of the KK zero mode is localized around the wall only for 
the bulk scalar field case.  For the KK zero mode ansatz (\ref{2ndkkans}), 
the bulk $p$-form potential is localized around $y=0$, spread evenly in 
the bulk and localized around $|y|=1/K$, when $p>(D-1)/2$, $p=(D-1)/2$ 
and $p<(D-1)/2$, respectively, as can be seen from the integrand of the 
$y$-integration in Eq. (\ref{kkact2}).  So, for $D=5$, the distribution 
of the KK zero mode is localized around the wall only for the bulk 
3-form potential case.  (Also, with such criterion, the KK zero mode for 
the bulk graviton is localized around both dilatonic and non-dilatonic 
domain walls.)  

Second, we consider the dimensional reduction of a bulk $p$-form potential 
to a $(p-1)$-form potential.  The KK ansatz is chosen to be
\begin{equation}
A_{\mu_1\dots\mu_{p-1}y}(x^{\mu},y)=a_{\mu_1\dots\mu_{p-1}}(x^{\mu})v(y),
\label{kkansm}
\end{equation}
with all the other components taken to vanish.  Substituting this ansatz 
into Eq. (\ref{peqn}), we obtain the following equation satisfied by 
$v(y)$:
\begin{equation}
\partial_y\left[{\cal W}^{{D-2p-1}\over 2}v\right]=0,
\label{eqforv}
\end{equation}
from which we obtain $v\sim{\cal W}^{-{{D-2p-1}\over 2}}$.  So, the KK 
ansatz (\ref{kkansm}) coincides with the ansatz (\ref{2ndkkans}) proposed 
in Ref. \cite{dl}, thereby providing a justification for such choice.  

It is pointed out in Ref. \cite{dl} that if we would additionally require 
the consistency with the bulk Einstein's equations then only the $p=0$ 
case with the ansatz (\ref{pformpot}) and the $p=D-2$ case with the ansatz 
(\ref{2ndkkans}) are allowed.  Actually, considering the Einstein's 
equations means looking for the solution for gravitating (non-dilatonic) 
charged $(p-1)$-brane or $(p-2)$-brane (coupled to the bulk $p$-form 
potential) within the domain wall, depending on the choice of the KK zero 
mode ansatz.  We have seen in our previous works \cite{youm4,youm5} that 
such solutions do not always exist due to the constraint following 
from the equations of motion.  In this paper and other related papers, where 
just the KK modes of bulk fields are studied, the gravitational backreaction 
of the bulk fields are ignored (thereby, the bulk metric remaining as a 
static background in which the bulk fields propagate) and therefore the 
Einstein's equations should not be considered.  If we would insist on 
consistency with the Einstein's equations, then we should rather consider 
the solution for (non-dilatonic) charged brane within the wall, which, 
according to Ref. \cite{youm4}, does not exist except for the $p=0,D-2$ 
cases mentioned in the above, as the bulk metric, instead of the domain 
wall solution (\ref{dildwsol}).   In the case in which the kinetic term 
for the bulk form potential has the dilaton factor $e^{2a_p\phi}$, the 
solution for the charged brane exists for any $p$, provided $a$ and $a_p$ 
satisfy the constraint resulting from the equations of motion 
\cite{youm4,youm5}.

\end{document}